\documentclass[preprintnumbers, aps, floatfix, onecolumn,
preprintnumbers, letterpaper, superscriptaddress,nofootinbib,11pt]{revtex4}
\usepackage{dcolumn}
\usepackage{graphicx}
\usepackage{amsmath}
\usepackage{amsfonts}
\usepackage{amssymb}
\usepackage{psfrag}
\usepackage{wrapfig}
\usepackage{subfigure}
\usepackage{makeidx}
\usepackage{bm}
\usepackage{epsf}
\usepackage[colorlinks=true,citecolor=blue,linkcolor=black,urlcolor=black]{hyperref}
\usepackage{color}
\usepackage{multirow,dcolumn}
\usepackage{graphicx}

\graphicspath{{Images/}}

\begin{document}

\title{Thermodynamical and dynamical properties of Charged BTZ Black Holes}
\author{Zi-Yu Tang}
\email{tangziyu@sjtu.edu.cn}
\affiliation{Center for Astronomy and Astrophysics, Department of Physics and Astronomy,
Shanghai Jiao Tong University, Shanghai 200240, China}
\author{Cheng-Yong Zhang}
\email{zhangcy0710@pku.edu.cn}
\affiliation{Center for High-Energy Physics, Peking University, Beijing 100871, China}
\author{Mahdi Kord Zangeneh}
\email{mkzangeneh@shirazu.ac.ir}
\affiliation{Physics Department, Faculty of Science, Shahid Chamran University of Ahvaz
61357-43135, Iran}
\affiliation{Research Institute for Astronomy and Astrophysics of Maragha
(RIAAM)-Maragha, IRAN, P. O. Box: 55134-441}
\affiliation{Center for Astronomy and Astrophysics, Department of Physics and Astronomy,
Shanghai Jiao Tong University, Shanghai 200240, China}
\affiliation{Physics Department and Biruni Observatory, College of Sciences, Shiraz
University, Shiraz 71454, Iran}
\author{Bin Wang}
\email{wang\_b@sjtu.edu.cn}
\affiliation{Center for Astronomy and Astrophysics, Department of Physics and Astronomy,
Shanghai Jiao Tong University, Shanghai 200240, China}
\author{Joel Saavedra}
\email{joel.saavedra@pucv.cl}
\affiliation{Instituto de Fisica, Pontificia Universidad Catolica de Valparaiso, Casilla
4950, Valparaiso, Chile}

\begin{abstract}
We investigate the spacetime properties of BTZ black holes in the presence
of the Maxwell field and Born-Infeld field and find rich properties in the
spacetime structures when the model parameters are varied. Employing the
Landau-Lifshitz theory, we examine the thermodynamical phase transition in
the charged BTZ black holes. We further study the dynamical perturbation in
the background of the charged BTZ black holes and find different properties
in the dynamics when the thermodynamical phase transition occurs.
\end{abstract}

\maketitle


\section{Introduction}

The discovery of three dimensional black holes by Ba$\mathrm{\tilde{n}}$%
ados, Teitelboim and Zanelli (BTZ black holes) \cite{Banados:1992wn} has
been known as one of the greatest achievements in the study of gravity. The
BTZ black hole provides us a framework to understand gravitational
interactions in low dimensional spacetimes \cite{Witten:2007kt}. It also
accommodates a simpler setting to explore many of the mysteries of black
hole statistical mechanics in higher dimensions \cite{Carlip}. In three
spacetime dimensions, general relativity becomes a topological field theory,
whose dynamics can be largely described holographically by a two dimensional
conformal field theory at the boundary of spacetime \cite{Carlip}. Thus the
BTZ black hole is a natural environment to realize the idea of AdS/CFT
correspondence. It was found that quasinormal modes determining the
relaxation time of rotating BTZ black hole perturbations are exactly in
agreement with the location of the poles of the retarded correlation
function of the corresponding perturbations in the dual conformal field
theory \cite{Birmingham}. This serves as a quantitative test of the AdS/CFT
correspondence, see \cite{Wang:2005vs,Konoplya:2011qq} for a review and
references therein on this topic.

Considerable progresses have been made for the study of BTZ black holes
where gravity minimally coupled to matter fields \cite{Carlip95,Carlip}.
When the 2+1 gravity is coupled to electromagnetism, we have the charged BTZ
black hole solution \cite{Banados:1992wn}. But compared with the neutral
holes, the charged BTZ black hole has not been thoroughly investigated. One
of the possible reasons is that the charged BTZ black hole is not a
spacetime of constant curvature \cite{BTZ94}. This makes the analysis in
terms of identifications in AdS space not applicable for charged BTZ black
holes \cite{BTZ94}, which becomes an obstacle to understand the geometrical
properties there. Another reason is that there is a logarithmic function in
the metric expression of the charged BTZ black hole, which makes the
analytic investigation difficult.

The motivation of this work is to study carefully the charged BTZ black
holes and present clearly the physical properties once enveiled by
mathematical difficulty. We will first concentrate on the charged BTZ black
hole solutions obtained from the standard Einstein-Maxwell equations in 2+1
spacetime dimensions with a negative cosmological constant. Then we will
generalize our discussions to charged BTZ black holes obtained when the
nonlinear Born-Infeld (BI) electromagnetism is brought in \cite%
{Mansoori:2015sit,Hendi:2015wxa}. The Born-Infeld theory (BI theory) was
first introduced to solve infinite self-energy problem by imposing a maximum
strength of the electromagnetic field \cite{Born:1934gh}. The generalization
of the Maxwell field to the nonlinear electrodynamics will lead to new
solutions and introduce new phenomena to the system under consideration \cite%
{Born:1934gh}. It is of great interest to investigate how the highly
nonlinear corrections to the gauge matter fields influence the bulk black
hole spacetime structure and its dynamical and thermodynamical properties in 
$2+1$ gravity.

In this paper we will first go over charged BTZ black hole solutions with
standard Maxwell field and nonlinear BI field. We will discuss the spacetime
structures, and disclose rich spacetime properties of the black hole with
the presence of the charge parameter. The strength of the charge parameter
determines whether the phase transition can happen or not. We will further
use the Landau-Lifshitz theory for thermodynamic fluctuations to discuss the
thermodynamical phase transitions in the charged BTZ black holes. We will
show that some second moments diverge in the extreme limit, indicating that
thermodynamical phase transition occurs.

It was argued that at the phase transition point, when the nonextreme black
hole becomes extreme, the Hawking temperature is zero which indicates that
for the extreme black hole there is only superradiation but no Hawking
radiation, which is in sharp difference from that of the nonextreme black
holes. Different radiation properties between extreme and nonextreme black
holes were used as an indication of the occurrence of the second order phase
transition \cite{Pavon,Pavon:1991kh,Cai:1993aa}.

However, there are some other discussions on Hawking radiation when Hawking
temperature is zero \cite{Chen:2012zn,Ong:2014nha}. For sufficiently low
black hole temperature, Hiscock and Weems modeled the evaporation of
4-dimensional asymptotically flat charged black hole by $\dfrac{dM}{dt}%
=-\alpha a T^{4}\sigma+\dfrac{Q}{r_{h}}\dfrac{dQ}{dt}$. They emphasized that
charged particles emission can be modeled separately from the thermal
Hawking flux of neutral particles, but they are all part of Hawking
radiation. This is because emission of charged particle is thermodynamically
related to a chemical potential associated with the electromagnetic field of
the black hole. Even in the limit $T=0$, the mass loss occurs from $\dfrac{dQ%
}{dt}$ term alone due to pair production by the Schwinger effect. The
presence of particle production at zero temperature can also be read from
Hawking's original formula $\langle N_{j\omega lp} \rangle=\dfrac{%
\Gamma_{j\omega lp}}{exp((\omega-e\Phi)/T)\pm 1}$ of \cite{Hawking:1974sw}
for the number of particle emission. See also Gibbons \cite{Gibbons:1975kk}
for discussions regarding emission from a charged black hole.

These subtleties regarding particle emission when Hawking temperature is
zero make the classification of different phases in nonextreme and extreme
black holes through different radiation properties somewhat subtle and
difficult. Then there comes a question whether there are some other
phenomena to indicate the sharp difference when the phase transition occurs
between the nonextreme and extreme black holes. This serves as another
motivation of the present paper. We will focus on the dynamical properties
of the charged BTZ black holes by studying the quasinormal modes of scalar
perturbation. We find that QNMs can serve as another probe for the phase
transition between the nonextreme and extreme black holes. The results tell
us that the extreme charged BTZ black hole is easily destroyed if we add
more perturbations to the system, while the nonlinearity of the
electromagnetic field can protect the black hole spacetime partially and
make the perturbation outside the black hole decay faster.

\section{ Charged BTZ black hole solutions}

In this section, we first review the derivation of charged BTZ black hole
solutions in the presence of linear Maxwell (LM) \cite{Banados:1992wn} and
nonlinear Born-Infeld (BI) \cite{Mansoori:2015sit} electrodynamics. Then, we
discuss the spacetime properties of these charged BTZ black holes.

Let us begin with the action of Einstein gravity in the presence of gauge
field,%
\begin{equation}
S_{grav}=S_{EH}+S_{gauge}+S_{GH},  \label{Action}
\end{equation}%
where $S_{EH}$, $S_{gauge}$ and $S_{GH}$ are the Einstein-Hilbert, gauge
field and Gibbons-Hawking actions defined as 
\begin{eqnarray}
S_{EH} &=&-\frac{1}{16\pi }\int_{M}d^{3}x\sqrt{-g}\left( R+\frac{2}{l^{2}}%
\right) , \\
S_{gauge} &=&-\frac{1}{16\pi }\int_{M}d^{3}x\sqrt{-g}L(F), \\
S_{GH} &=&-\frac{1}{8\pi }\int_{\partial M}d^{2}x\sqrt{-h}K,
\end{eqnarray}%
in which $R$ is the Ricci scalar for the bulk manifold $M$, $l$ is the AdS
radius and $L(F)$ is the Lagrangian of electrodynamic field $F_{\mu \nu }$
where $F=F_{\mu \nu }F^{\mu \nu }$, $F_{\mu \nu }=\partial _{\lbrack \mu
}A_{\nu ]}$ and $A_{\nu }$ is the electrodynamic gauge potential. $h$ is the
determinant of the $2$-dimensional metric on the boundary of manifold $M$ ($%
\partial M$) and $K$ is the trace of the extrinsic curvature of the
boundary. Lagrangians of electrodynamic field for linear Maxwell and
nonlinear BI cases are%
\begin{equation}
L(F)=\left\{ 
\begin{array}{ll}
-F & \text{LM} \\ 
4\beta ^{2}\left( 1-\sqrt{1+\frac{F}{2\beta ^{2}}}\right) & \text{BI}%
\end{array}%
\right. ,
\end{equation}%
where $\beta $ is the parameter of nonlinearity. Nonlinear BI
electrodynamics reduces to the linear Maxwell case when $\beta\rightarrow
\infty $. Varying the action (\ref{Action}) with respect to the metric $%
g_{\mu \nu }$ and the gauge potential $A_{\mu }$, we obtain%
\begin{eqnarray}
R_{\mu \nu }-\frac{1}{2}g_{\mu \nu }\left( R+\frac{2}{l^{2}}\right) &=&\frac{%
1}{2}g_{\mu \nu }L(F)-2F_{\mu \sigma }F_{\nu }^{\sigma }\frac{dL(F)}{dF},
\label{EiE} \\
\partial _{\mu }\left( \sqrt{-g}\frac{dL(F)}{dF}F^{\mu \nu }\right) &=&0.
\label{ElE}
\end{eqnarray}%
Substituting the gauge potential $A_{\mu }=\varphi (r)\delta _{\mu }^{0}$
into Eq. (\ref{ElE}), we obtain the nonvanishing component of the
electrodynamic field tensor 
\begin{equation}
F_{tr}=-F_{rt}=\frac{q}{r}\times \left\{ 
\begin{array}{ll}
1 & \text{LM} \\ 
\Gamma ^{-1} & \text{BI}%
\end{array}%
\right. ,
\end{equation}%
where $q$ is a constant related to the total charge of the black hole and $%
\Gamma =\sqrt{1+q^{2}/\left( r^{2}\beta ^{2}\right) }$. Eq. (\ref{EiE})
admits the charged BTZ black hole solution as%
\begin{equation}
ds^{2}=-f(r)dt^{2}+\frac{dr^{2}}{f(r)}+r^{2}d\theta ^{2},
\end{equation}%
in which 
\begin{equation}
f(r)=\frac{r^{2}}{l^{2}}-m+\left\{ 
\begin{array}{ll}
-2q^{2}\ln {\frac{r}{l}} & \text{LM} \\ 
2r^{2}\beta ^{2}\left( 1-\Gamma \right) +q^{2}\left[ 1-2\ln \left( r\frac{%
\left( 1+\Gamma \right) }{2l}\right) \right] \  & \text{BI}%
\end{array}%
\right. ,  \label{metric}
\end{equation}%
where $m$ is a constant which is proportional to the total mass of the black
hole and can be obtained by using the fact that metric function vanishes at
the event horizon $r_{+}$%
\begin{equation}
m=\left\{ 
\begin{array}{ll}
\frac{r_{+}^{2}}{l^{2}}-2q^{2}\ln \left( \frac{r_{+}}{l}\right) & \text{LM}
\\ 
\frac{r_{+}^{2}}{l^{2}}+2r_{+}^{2}\beta ^{2}\left( 1-\Gamma _{+}\right)
+q^{2}\left[ 1-2\ln \left( r_{+}\frac{\left( 1+\Gamma _{+}\right) }{2l}%
\right) \right] & \text{BI}%
\end{array}%
\right. ,  \label{m}
\end{equation}%
where $\Gamma _{+}=\sqrt{1+q^{2}/\left( r_{+}^{2}\beta ^{2}\right) }$.

In the following, we will explore the spacetime properties of linearly and
nonlinearly charged BTZ black holes.

\subsection{Linearly charged BTZ black holes}

\begin{figure}[t]
\centering
\includegraphics[width=0.45\textwidth]{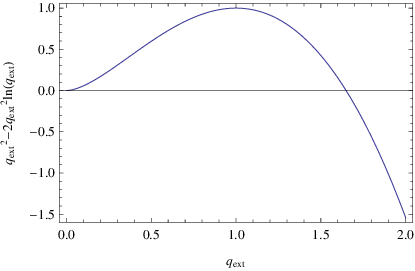}
\caption{{}The behavior of the left hand side of Eq.(\protect\ref{eq:q})
versus $q$. There is a maximum at $q_{\mathrm{ext}}=1$. }
\label{fig:q}
\end{figure}

The metric function for the linearly charged BTZ black hole reads%
\begin{equation}
f(r)=\frac{r^{2}}{l^{2}}-m-2q^{2}\ln {\frac{r}{l}}.
\end{equation}%
In \cite{Hendi:2015wxa}, it was argued that for fixed parameters $l=1$ and $%
q=1$, there is a naked singularity when $m$ is sufficiently small, while the
mass parameter is at a critical value, there is an extreme BTZ black hole.
When the mass is above the critical value, the charged BTZ black hole can
have two horizons enveloping the central singularity, namely the inner
Cauchy horizon $r_{-}$ and the outer event horizon $r_{+}$.

How about the situation if we treat $q$ as a variable? The answer was not
provided in \cite{Hendi:2015wxa}. In order to study the extreme charged BTZ
black hole, we first let the metric function $f(r)$ to be zero which can
tell us the location of the horizon. Furthermore we require the derivative
of the metric function $f^{\prime }(r)$ to vanish at the horizon which
ensures the extremal condition to be satisfied. From these conditions, 
\begin{equation}
f(r_{+\mathrm{ext}})=\frac{r_{+\mathrm{ext}}^{2}}{l^{2}}-m_{\mathrm{ext}%
}-2q_{\mathrm{ext}}^{2}\ln {\frac{r_{+\mathrm{ext}}}{l}}=0,\text{ \ and \ }%
f^{\prime }(r_{+\mathrm{ext}})=\frac{2r_{+\mathrm{ext}}}{l^{2}}-\frac{2q_{%
\mathrm{ext}}^{2}}{r_{+\mathrm{ext}}}=0,
\end{equation}
we can have an equation relating $q_{\mathrm{ext}}$ and $m_{\mathrm{ext}}$ 
\begin{equation}
q_{\mathrm{ext}}^{2}-2q_{\mathrm{ext}}^{2}\ln {q}_{\mathrm{ext}}=m_{\mathrm{%
ext}}.  \label{eq:q}
\end{equation}%
The behavior of the left-hand-side of Eq. (\ref{eq:q}) is shown in Fig. \ref%
{fig:q}, which has a peak equaling to one when $q=1$. When $m=1$, there is
only one solution of Eq. (\ref{eq:q}), which is at $q=1$ and the charged BTZ
black hole is extreme. When $0<m<1$, there will be two solutions for $q_{%
\mathrm{ext}}$ from Eq. (\ref{eq:q}). The black hole can be extreme for
these two charge values when $m<1$. However when $m>1$, there is no solution
of (\ref{eq:q}), which indicates that the extreme black hole condition
cannot be respected so that the charged BTZ black hole is always nonextreme
when $m>1$.

\begin{figure}[t]
\centering%
\subfigure[~$m=0.5$]{
 \label{fig2a}\includegraphics[width=.3\textwidth]{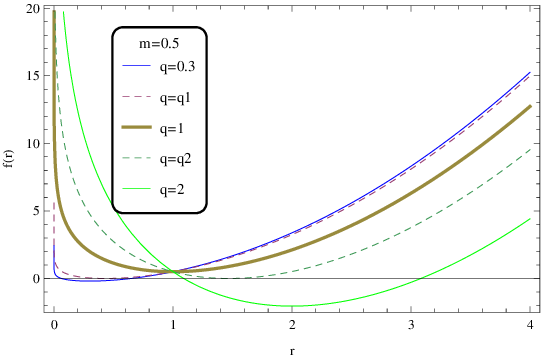} } 
\subfigure[~$m=1$]{
 \label{fig2b}\includegraphics[width=.3\textwidth]{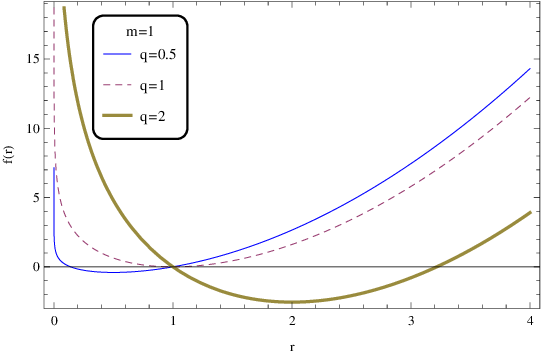} } 
\subfigure[~$m=2$]{
 \label{fig2c}\includegraphics[width=.3\textwidth]{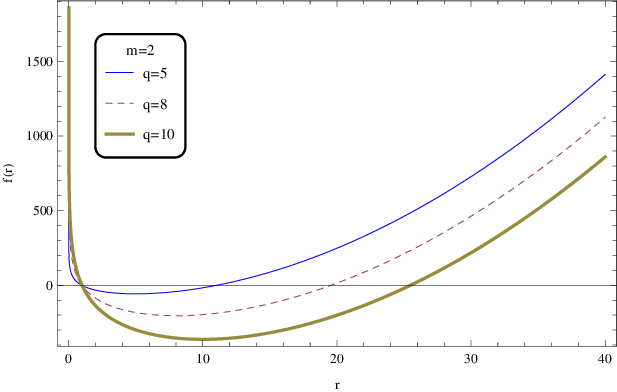}}
\caption{{}The behaviors of $f(r)$ for different values of $q$ when $m=0.5$, 
$1$ and $2$.}
\label{fig:MF}
\end{figure}

To have a clearer picture, we plot the metric function $f(r)$ in Fig. \ref%
{fig:MF} for different values of $m$. For $m<1$ (we choose $m=0.5$), we have
two values $q_{1}$ and $q_{2}$ to accommodate the extreme black hole. We can
see From Fig. \ref{fig2a} that when $q<q_{1}$ and $q>q_2$, the black hole is
nonextreme with two horizons. But in the range $q_{1}<q<q_{2}$, there is no
root of the metric function so that the black hole does not exist. When $m=1$%
, we see that there is just one value for the critical charge, namely $q=1$,
to accommodate the extreme charged BTZ black holes. Below or above this
critical charge, the black hole is always nonextreme, which is shown in Fig. %
\ref{fig2b}. For $m>1$, it is clear from Fig. \ref{fig2c} that there are
always two horizons enveloping the central singularity and the black hole is
nonextreme. 
\begin{figure}[t]
\centering%
\subfigure[~$R$ vs $q$]{
 \label{fig3a}\includegraphics[width=.45\textwidth]{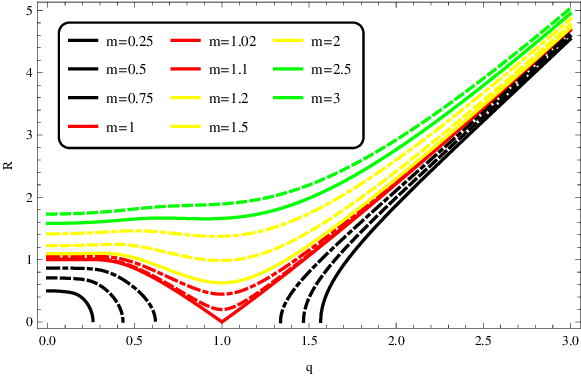} } 
\subfigure[~$T$ vs $q$]{
 \label{fig3b}\includegraphics[width=.45\textwidth]{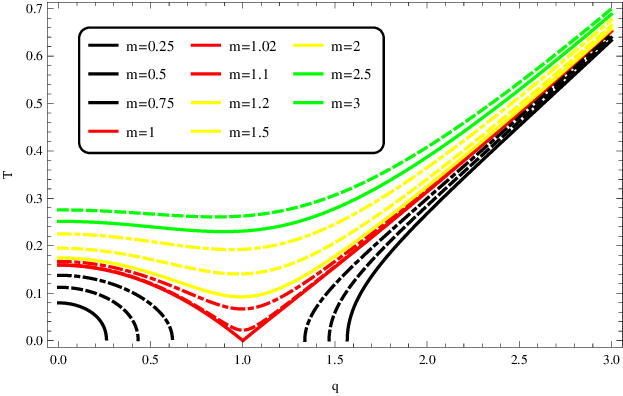} }
\caption{{} The behaviors of distance $R$ between inner and outer horizons
(left) and the temperature $T$ (right) of charged BTZ black hole.}
\label{fig:RT-q}
\end{figure}

We further define the coordinate difference $R=r_{+}-r_{-}$ between the
inner and outer horizons to show its dependence on the parameter $q$ for
different chosen mass values in Fig. \ref{fig3a}. When the black hole
geometrical mass is smaller than one ($m<1$), there will be two horizons
protecting the central singularity when the black hole charge is small.
These two horizons approach each other with the increase of the black hole
charge and the nonextreme black hole becomes extreme at $q=q_{1}$. When the
charge parameter is between $q_{1}$ and $q_{2}$, there is no root for the
metric function $f(r)$ and the black hole does not exist. At $q=q_{2}$ for
the chosen $m$ value smaller than one, there is only one black hole horizon
to appear to envelop the singularity again. The difference between event and
Cauchy horizons becomes bigger when the black hole is more charged when $%
q>q_2$. When $m$ approaches one from below, we see from Fig. \ref{fig3a}
that $q_1$ and $q_2$ approach each other. They finally merge when $m=1$ and
only one extreme charged BTZ black hole exists for $q=1$. Below $q=1$, with
the increase of the electric charge, the two horizons become closer to each
other, degenerate finally at $q=1$. But when above $q=1$, the horizons can
separate again when $q$ grows and the black hole becomes more nonextreme. In
the case when $m>1$, the two horizons will never degenerate so that the
black hole is always nonextreme. There exists a turning point of the mass
value $m_{r}$. In the range $1<m<m_{r}$, the two horizons can come closer to
each other and then separate away with the increase of $q$, while when $%
m>m_{r}$, the distance between the two horizons of the black hole will
increase monotonically.


We can calculate the charged BTZ black hole temperature%
\begin{equation}
T=\frac{f^{\prime }\left( r_{+}\right) }{4\pi }=\frac{1}{2\pi }\left( \frac{%
r_{+}}{L^{2}}-\frac{q^{2}}{r_{+}}\right)
\end{equation}%
which is shown in Fig. \ref{fig3b}. The black hole temperature confirms the
property we disclosed in the spacetime structure. When the black hole mass
is small ($m<1$), the black hole temperature decreases to zero with the
increase of the black hole charge from $0$ to $q_{1}$. When the charge is
big enough ($q\geq q_{2}$), the black hole can exist and the temperature
will rise from zero. When $m=1$, the temperature decreases to zero at $q=1$
and then rises  with the increase of charge $q$. For $m>1$, we see from Fig. %
\ref{fig3b} that there exists a turning point of the mass $m_{t}$, when  $%
1<m<m_{t}$, with the increase of the electric charge, the temperature will
first decrease and then increase but the minimum of the temperature is
always above zero. When $m>m_{t}$, the temperature increases monotonically
with the increase of the electric charge.

We can clearly see that the qualitative behavior of $T$ is closely related
to $R$.

\subsection{Nonlinearly charged BTZ black holes}

The metric function of nonlinearly BI charged BTZ black hole is 
\begin{equation}
f(r)=\frac{r^{2}}{l^{2}}-m+2r^{2}\beta ^{2}\left( 1-\Gamma \right) +q^{2}%
\left[ 1-2\ln \left( r\frac{\left( 1+\Gamma \right) }{2l}\right) \right] .
\end{equation}%
When $\beta\rightarrow \infty $, the results reduce to that presented in the
above subsection. When $\beta $ becomes smaller, the nonlinearity becomes
stronger.

We repeat the discussion above to examine the spacetime structure for the
nonlinearly charged BTZ black hole. The location of the horizon and the
extreme condition are described by 
\begin{equation}
f(r_{+})=\frac{r_{+}^{2}}{l^{2}}-m_{\mathrm{ext}}+2r_{+}^{2}\beta ^{2}\left(
1-\Gamma _{+}\right) +q_{\mathrm{ext}}^{2}\left[ 1-2\ln \left( r_{+}\frac{%
\left( 1+\Gamma _{+}\right) }{2l}\right) \right] =0,
\end{equation}%
and%
\begin{equation}
f^{\prime }(r_{+})=\frac{2r_{+}^{2}\beta ^{2}\left( 1+\Gamma _{+}\right)
+2q_{\mathrm{ext}}^{2}\left( 1-2l^{2}\beta ^{2}\Gamma _{+}\right) }{%
l^{2}\beta ^{2}r_{+}\left( 1+\Gamma _{+}\right) \Gamma _{+}}=0,
\end{equation}%
where we can find the relation between $q_{\mathrm{ext}}$ and $m_{\mathrm{ext%
}}$ in the form 
\begin{equation}
q_{\mathrm{ext}}^{2}-2q_{\mathrm{ext}}^{2}\ln {\frac{q_{\mathrm{ext}}}{%
2L\beta }}\sqrt{1+4l^{2}\beta ^{2}}=m_{\mathrm{ext}}.  \label{eq:q2}
\end{equation}

\begin{figure}[t]
\centering
\includegraphics[width=0.45\textwidth]{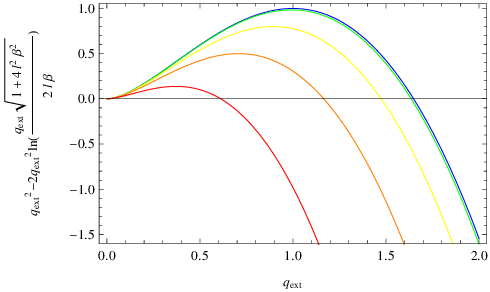}
\caption{{}The behaviors of the left hand side of Eq. (\protect\ref{eq:q2})
versus $q$ for $\protect\beta =0.2$, $0.5$, $1$, $4$, and infinity
respectively from below to the up lines, where we take $l=1$. }
\label{fig:qNL}
\end{figure}

From Fig. \ref{fig:qNL}, we see that there always exists a maximum value for
the left hand side of Eq. (\ref{eq:q2}). This maximum value $4l^{2}\beta
^{2}/\left( 1+4l^{2}\beta ^{2}\right) $ happens at $q_{\mathrm{ext}}=2l\beta
/\sqrt{1+4l^{2}\beta ^{2}}$. When $m=4l^{2}\beta ^{2}/\left( 1+4l^{2}\beta
^{2}\right) $, there is only one solution $q_{1}$ which can make the black
hole extreme. For $m<4l^{2}\beta ^{2}/\left( 1+4l^{2}\beta ^{2}\right) $,
there are two solutions $q_{1}$ and $q_{2}$ which satisfy the extreme black
hole condition. Interestingly, $q_{1}$ and $q_{2}$ can become closer when
the nonlinearity is increased in the electromagnetic field. When $%
m>4l^{2}\beta ^{2}/\left( 1+4l^{2}\beta ^{2}\right) $, the extreme condition
cannot be satisfied so that the black hole is always nonextreme.

\begin{figure}[t]
\centering%
\subfigure[~$m=0.25$]{
 \label{fig4a}\includegraphics[width=.3\textwidth]{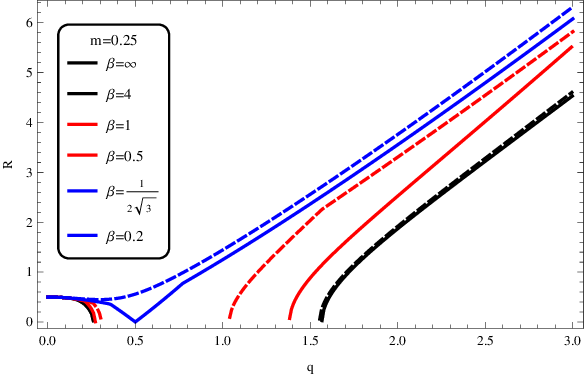} } 
\subfigure[~$m=1$]{
 \label{fig4b}\includegraphics[width=.3\textwidth]{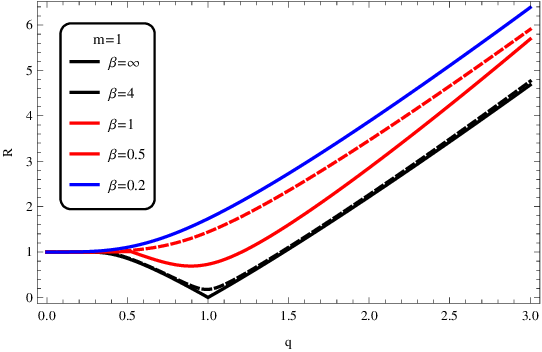} } 
\subfigure[~$m=2$]{
 \label{fig4c}\includegraphics[width=.3\textwidth]{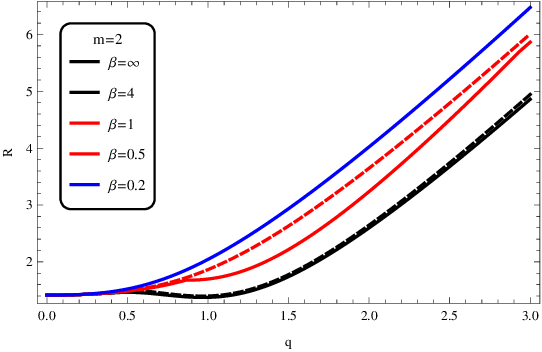} }
\caption{{}The distance $R$ between inner and outer horizons for $m=0.25$, $%
1 $ and $2$ for nonlinearly charged BTZ black holes.}
\label{fig:R-qNL}
\end{figure}

To display the property more clearly, we again plot in Fig. \ref{fig:R-qNL}
the behavior of the distance between the inner and the outer horizons $R$
with respect to $q$ for $m=0.25$, $1 $ and $2$, respectively. For $m<1$ ($%
m=0.25$), we find that with the decrease of $\beta$, the two extreme values
of the charge $q$ get closer to each other, become one and then disappear.
This means that when the nonlinearity in the electromagnetic field is strong
enough, there is no extreme charged BTZ black hole. For $m=1$, the extreme
black hole only exists when the gravity is coupled to the standard Maxwell
field. For gravity coupled to the BI field, when the nonlinearity is not
strong enough, the two horizons of the black hole can approach to each other
first and then separate with the increase of the electric charge. But they
cannot merge. When the electromagnetic field is strongly nonlinear, the
difference between two horizons always becomes bigger with the increase of
the black hole charge. For $m>1$, the influence of the nonlinearity is
similar to that described for $m=1$ case.

\begin{figure}[t]
\centering%
\subfigure[~$m=0.25$]{
 \label{fig5a}\includegraphics[width=.3\textwidth]{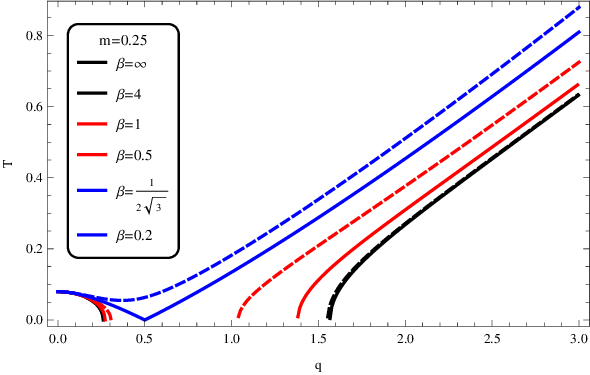} } 
\subfigure[~$m=1$]{
 \label{fig5b}\includegraphics[width=.3\textwidth]{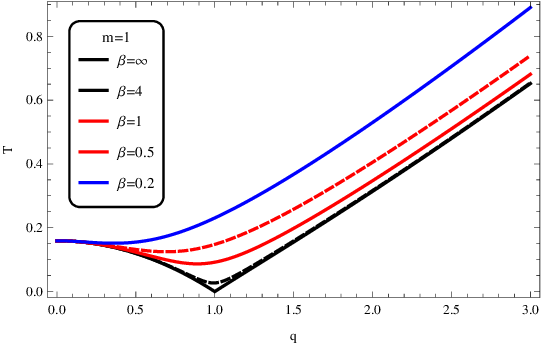} } 
\subfigure[~$m=2$]{
 \label{fig5c}\includegraphics[width=.3\textwidth]{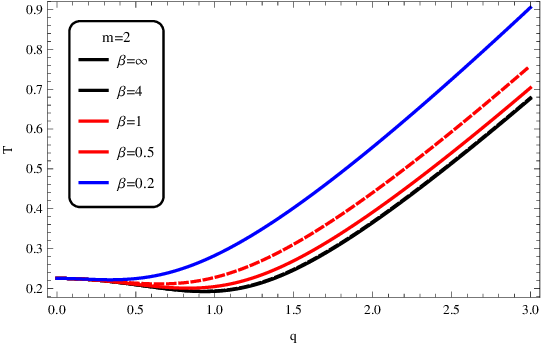} }
\caption{{}The behavior of Hawking temperature $T$ for $m=0.25$, $1$ and $2$
for nonlinearly charged black holes.}
\label{fig:T-qNL}
\end{figure}

The behavior of Hawking temperature $T$ for the nonlinearly charged BTZ
black holes is shown in Fig. \ref{fig:T-qNL}. It gives us qualitatively the
similar behavior as we discussed for $R$. Compared with the linearly charged
BTZ black hole, the nonlinearity in the electromagnetic field makes the
Hawking temperature higher and it is more difficult to reach zero
temperature in the nonlinearly charged BTZ black holes. 

\section{ Thermodynamical phase transitions}

Thermodynamics of the BTZ black holes have been discussed in \cite{BTZ94}.
More references can be found in reviews \cite{Carlip95, Carlip}. Most
studies were focused on the BTZ black holes where gravity minimally coupled
to matter fields. For the charged BTZ black holes, the first law of
thermodynamics was constructed in \cite{alexis}, the entropy was examined in 
\cite{Cadoni,Myung,Wang2}, the mass bound and thermodynamical behavior was
investigated in \cite{Cadoni2010} and the phase transition was studied in 
\cite{Wang}. Recently the thermodynamical stability of the charged BTZ black
hole was investigated in \cite{Hendi:2015wxa} by the method of examining the
heat capacity first proposed by Davies \cite%
{Davies:1978mf,Davies:1978zz,Davies:1989ey}. The extension was performed to
the nonlinearly charged BTZ black holes \cite{Hendi2}.

Here we are going to examine the thermodynamical phase transition in the
charged BTZ black holes more carefully. It was argued in \cite{Pavon} that
the heat capacity cannot be the true character to mark the phase transition,
because there is no sharp change in physical properties at the point of the
phase transition obtained from the heat capacity.

Besides, the comparison of the free energy of two competing configurations are often used in the context of first order phase transition, such as the Hawking-Page phase transition, which describes the competition between a pure anti-de Sitter spacetime and an AdS black hole spacetime. However, the phase transition we studied is between non-extremal and extremal black holes. This is a second order phase transition. So we turn to use the Landau-Lifshitz theory of thermodynamical perturbations. Employing the
Landau-Lifshitz theory of thermodynamic fluctuations, the fluctuations in
the rate of change of mass, angular momentum and other relevant quantities
of different black holes were examined and it was found that some second
moments in the fluctuation of relevant quantities diverge when the black
holes become extreme, which marks the occurrence of the second order phase
transition. Below we will further use the Landau-Lifshitz theory to study
the phase transition in the charged BTZ black holes.

According to Landau-Lifshitz theory \cite{LL1}, in a fluctuation-dissipative
process, the flux $\dot{X}_{i}$ (dot shows temporal derivative) of a given
thermodynamic quantity $X_{i}$ is expressed by%
\begin{equation}
\dot{X}_{i}=-\sum_{j}\Gamma _{ij}\chi _{j},
\end{equation}%
where $\Gamma _{ij}$ and $\chi _{i}$ are the phenomenological transport
coefficients and thermodynamic force conjugate to the flux $\dot{X}_{i}$,
respectively. Also, the entropy production rate is given by%
\begin{equation}
\dot{S}=\sum_{i}\pm \chi _{i}\dot{X}_{i},
\end{equation}%
The second moments in the fluctuations of the fluxes read%
\begin{equation}
\left\langle \delta \dot{X}_{i}\delta \dot{X}_{j}\right\rangle =\left(
\Gamma _{ij}+\Gamma _{ji}\right) \delta _{ij},  \label{smom}
\end{equation}%
where the angular brackets denote the mean value with respect to the steady
state and the fluctuations $\delta \dot{X}_{i}$ are the spontaneous
deviations from the steady state value $\left\langle \dot{X}%
_{i}\right\rangle $ (we set $k_{B}=1$). The Kronecker $\delta _{ij}$ in Eq. (%
\ref{smom}) is to guarantee that correlations vanish when two fluxes are
independent. We can obtain the rate of entropy production as%
\begin{eqnarray}
\dot{S}\left( M,Q\right) &=&\left( \frac{\partial S}{\partial M}\right) _{Q}%
\dot{M}+\left( \frac{\partial S}{\partial Q}\right) _{M}\dot{Q}  \notag \\
&=&\frac{\dot{M}}{T}-\frac{\dot{Q}}{T}\left( \frac{\partial M}{\partial Q}%
\right) _{S}  \notag \\
&=&\frac{\dot{M}}{T}-\frac{U\dot{Q}}{T},
\end{eqnarray}%
where we have used%
\begin{equation}
T=\left( \frac{\partial M}{\partial S}\right) _{Q},\text{ \ \ \ \ }U=\left( 
\frac{\partial M}{\partial Q}\right) _{S},  \label{TU}
\end{equation}%
and%
\begin{equation}
\left( \frac{\partial M}{\partial S}\right) _{Q}\left( \frac{\partial S}{%
\partial Q}\right) _{M}\left( \frac{\partial Q}{\partial M}\right) _{S}=-1,
\end{equation}%
in which $S$, $M$ and $Q$ are entropy, total mass and total charge of the
black hole defined as \cite{Hendi:2015wxa}%
\begin{equation}
S=\frac{\pi r_{+}}{2},\text{ \ \ }Q=\frac{q}{2}\text{ \ and \ }M=\frac{m}{8},
\label{SQM}
\end{equation}%
and $U$ is the electric potential energy. Using Eqs. (\ref{m}), (\ref{TU})
and (\ref{SQM}), one can calculate the temperature and the electric
potential energy as%
\begin{equation*}
T=\left\{ 
\begin{array}{ll}
\frac{S}{\pi ^{2}l^{2}}-\frac{Q^{2}}{S} & \text{LM} \\ 
\frac{S}{\pi ^{2}l^{2}}-\frac{Q^{2}}{S}+\frac{2S\beta ^{2}}{\pi ^{2}}\left(
1-\Gamma _{S}\right) +\frac{Q^{2}}{S\Gamma _{S}}+\frac{\pi ^{2}Q^{4}}{\beta
^{2}S^{3}\Gamma _{S}\left( 1+\Gamma _{S}\right) } & \text{BI}%
\end{array}%
\right. ,
\end{equation*}%
and%
\begin{equation*}
U=\left\{ 
\begin{array}{ll}
-2Q\ln \left( \frac{2S}{\pi l}\right) & \text{LM} \\ 
-2Q\ln \left( \frac{S}{\pi l}\left( 1+\Gamma _{S}\right) \right) +Q\left(
1-\Gamma _{S}^{-1}\right) -\frac{\pi ^{2}Q^{3}}{\beta ^{2}S^{2}\Gamma
_{S}\left( 1+\Gamma _{S}\right) } & \text{BI}%
\end{array}%
\right. ,
\end{equation*}%
where $\Gamma _{S}=\sqrt{1+Q^{2}\pi ^{2}/\left( S^{2}\beta ^{2}\right) }$.
It is worthwhile to mention that for the extreme black hole case where $%
f^{\prime }(r_{+})=0$, the Hawking temperature vanishes since $T=f^{\prime
}(r_{+})/4\pi $.

The rate of the mass loss is \cite{His,Cardoso:2005cd}%
\begin{equation}
\frac{dM}{dt}=-b\alpha \sigma T^{3}+U\frac{dQ}{dt}.  \label{Md}
\end{equation}%
The first term on the right side of (\ref{Md}) exhibits the thermal mass
loss due to Hawking radiation. Note that the power of the temperature is
dimension-dependent. This term is just the Stefan-Boltzmann law, in which $b$
denotes the radiation constant. The constant $\alpha $ is dependent on the
number of species of massless particles and the quantity $\sigma $ is the
geometrical optics cross-section. The second term on the right side of Eq. (%
\ref{Md}) is due to the mass loss of charged particles. In fact, It is the
same term that appears in the first law of black hole mechanics, namely $UdQ$%
.

According to what described and calculated above, we can obtain the
correlation functions or the second moments of the corresponding
thermodynamical quantities as%
\begin{equation}
\left\langle \delta \dot{M}\delta \dot{M}\right\rangle =-2T\dot{M},\text{ \
\ \ \ }\left\langle \delta \dot{Q}\delta \dot{Q}\right\rangle =-\frac{2T\dot{%
Q}}{U},\text{ \ \ \ \ }\left\langle \delta \dot{M}\delta \dot{Q}%
\right\rangle =U\left\langle \delta \dot{Q}\delta \dot{Q}\right\rangle ,
\end{equation}%
\begin{equation}
\left\langle \delta \dot{S}\delta \dot{M}\right\rangle =\frac{1}{T}%
\left\langle \delta \dot{M}\delta \dot{M}\right\rangle -\frac{U}{T}%
\left\langle \delta \dot{M}\delta \dot{Q}\right\rangle =\frac{1}{T}%
\left\langle \delta \dot{M}\delta \dot{M}\right\rangle -\frac{U^{2}}{T}%
\left\langle \delta \dot{Q}\delta \dot{Q}\right\rangle =-2\dot{M}+2U\dot{Q},
\end{equation}%
\begin{equation}
\left\langle \delta \dot{S}\delta \dot{Q}\right\rangle =\frac{1}{T}%
\left\langle \delta \dot{M}\delta \dot{Q}\right\rangle -\frac{U}{T}%
\left\langle \delta \dot{Q}\delta \dot{Q}\right\rangle =\frac{U}{T}%
\left\langle \delta \dot{Q}\delta \dot{Q}\right\rangle -\frac{U}{T}%
\left\langle \delta \dot{Q}\delta \dot{Q}\right\rangle =0
\end{equation}%
\begin{eqnarray}
\left\langle \delta \dot{S}\delta \dot{S}\right\rangle &=&\frac{1}{T^{2}}%
\left\langle \delta \dot{M}\delta \dot{M}\right\rangle +\frac{U^{2}}{T^{2}}%
\left\langle \delta \dot{Q}\delta \dot{Q}\right\rangle -\frac{2U}{T^{2}}%
\left\langle \delta \dot{M}\delta \dot{Q}\right\rangle =\frac{1}{T^{2}}%
\left\langle \delta \dot{M}\delta \dot{M}\right\rangle -\frac{U^{2}}{T^{2}}%
\left\langle \delta \dot{Q}\delta \dot{Q}\right\rangle  \notag \\
&=&-\frac{2\dot{M}}{T}+\frac{2U\dot{Q}}{T}=\frac{\left\langle \delta \dot{S}%
\delta \dot{M}\right\rangle }{T},
\end{eqnarray}%
\begin{equation}
\left\langle \delta \dot{T}\delta \dot{T}\right\rangle
=M_{SS}^{2}\left\langle \delta \dot{S}\delta \dot{S}\right\rangle
+M_{SQ}^{2}\left\langle \delta \dot{Q}\delta \dot{Q}\right\rangle
+2M_{SS}M_{SQ}\left\langle \delta \dot{S}\delta \dot{Q}\right\rangle
=M_{SS}^{2}\left\langle \delta \dot{S}\delta \dot{S}\right\rangle
+M_{SQ}^{2}\left\langle \delta \dot{Q}\delta \dot{Q}\right\rangle ,
\end{equation}%
\begin{equation}
\left\langle \delta \dot{S}\delta \dot{T}\right\rangle =\frac{M_{SS}}{T}%
\left\langle \delta \dot{S}\delta \dot{M}\right\rangle +\frac{M_{SQ}}{T}%
\left\langle \delta \dot{M}\delta \dot{Q}\right\rangle -\frac{UM_{SS}}{T}%
\left\langle \delta \dot{S}\delta \dot{Q}\right\rangle -\frac{UM_{SQ}}{T}%
\left\langle \delta \dot{Q}\delta \dot{Q}\right\rangle =M_{SS}\left\langle
\delta \dot{S}\delta \dot{S}\right\rangle ,
\end{equation}%
Note that, for calculating $\left\langle \delta \dot{T}\delta \dot{T}%
\right\rangle $ and $\left\langle \delta \dot{S}\delta \dot{T}\right\rangle $%
, we have used%
\begin{equation}
\dot{T}(S,Q)=\left( \frac{\partial T}{\partial S}\right) _{Q}\dot{S}+\left( 
\frac{\partial T}{\partial Q}\right) _{S}\dot{Q}=M_{SS}\dot{S}+M_{SQ}\dot{Q},
\end{equation}%
in which $M_{XY}=\partial ^{2}M/\partial X\partial Y$. Since $T$ vanishes
for the extreme black hole case, second moments $\left\langle \delta \dot{S}%
\delta \dot{S}\right\rangle $, $\left\langle \delta \dot{T}\delta \dot{T}%
\right\rangle $ and $\left\langle \delta \dot{S}\delta \dot{T}\right\rangle $
diverge there. The divergence of the second moment means that the
fluctuation is tremendous which breaks down the rigorous meaning of
thermodynamical quantities. This is just the characteristic of the point of
the phase transition. This property holds for the nonlinearly charged BTZ
black hole as well.

\section{The dynamical perturbations}

It is an interesting question how we can examine the phase transition
through physical phenomena. Hiscock and Weems's argument tells us that the
Hawking radiation is not exactly zero when the black hole temperature
vanishes \cite{Wang}. This makes the attempt to indicate of the second order
black hole phase transition through different radiation properties fade. In
this section we are going to explore the properties of the dynamical
perturbations in the charged BTZ black hole background. We will examine the
behavior of quasi normal modes (QNMs) of scalar perturbations, and will
argue that the dynamical perturbation behavior can serve as a new probe of
the phase transition between the extreme and nonextreme black holes.

QNM of black holes does not depend on perturbation fields outside black
holes. Instead, it only depends on properties of the background spacetime.
Thus QNMs can reflect the dynamical properties of black holes. It was argued
in \cite{Koutsoumbas:2006xj} that the QNMs of the electromagnetic
perturbations can give evidence of a second-order phase transition of a
topological black hole to a hairy configuration. Further supports that the
QNMs can be probes of the phase transition were provided in \cite%
{Shen:2007xk,Wang3}. In a recent paper, it was reported that the signature
of the Van der Waals like small-large charged AdS black hole phase
transition can also be observed in QNM \cite{Wang4}. It is of interest to
generalize the previous studies to investigate the dynamical perturbations
in the charged BTZ black holes and examine whether the QNMs of the
perturbation can again be a physical indication of the thermodynamical phase
transition.

In calculating the QNMs, we have tried three different numerical methods 
\cite{Konoplya:2011qq}, namely the Horowitz-Hobeny method, the Shooting
method and the Asymptotic iteration method (AIM). We found that the AIM is
the most precise and efficient method, especially when the black hole
approaches extremality. In the following we will introduce the AIM method
first and present the numerical results.

AIM was first used to solve eigenvalue problems of the second order
homogeneous linear differential equations \cite{Ciftci:2005xn}, and then
applied to the case of black hole QNMs. Considering a second order
homogeneous linear differential equation 
\begin{equation}
\chi ^{\prime \prime }=\lambda _{0}(x)\chi ^{\prime }+s_{0}(x)\chi ,
\label{eq:standardform}
\end{equation}%
where $\lambda _{0}(x)$ and $s_{0}(x)$ are smooth functions in some interval 
$[a,b]$ and taking the derivative of Eq.~(\ref{eq:standardform}) with
respect to $x$, we obtain the following equation 
\begin{equation}
\chi ^{\prime \prime \prime }=\lambda _{1}(x)\chi ^{\prime }+s_{1}(x)\chi ,
\end{equation}%
where 
\begin{equation}
\lambda _{1}(x)=\lambda _{0}^{\prime }(x)+s_{0}(x)+\lambda _{0}^{2}(x),\text{
\ and \ }s_{1}=s_{0}^{\prime }(x)+s_{0}(x)\lambda _{0}(x).
\end{equation}%
Repeating this step iteratively, we can obtain the $(n+2)$-th derivatives 
\begin{equation}
\chi ^{(n+2)}=\lambda _{n}(x)\chi ^{\prime }+s_{n}(x)\chi ,
\end{equation}%
where 
\begin{equation}
\lambda _{n}(x)=\lambda _{n-1}^{\prime }(x)+s_{n-1}(x)+\lambda
_{0}(x)\lambda _{n-1}(x),\text{ \ and \ }s_{n}(x)=s_{n-1}^{\prime
}(x)+s_{0}(x)\lambda _{n-1}(x).  \label{eq:AIM_iteration}
\end{equation}%
For sufficiently large $n$, the asymptotic aspect of the method is
introduced~as \cite{Cho:2012} 
\begin{equation}
\frac{s_{n}(x)}{\lambda _{n}(x)}=\frac{s_{n-1}(x)}{\lambda _{n-1}(x)}.
\label{asym}
\end{equation}%
The QNMs can be derived from the above ``quantization condition". However,
the derivatives of $\lambda _{n}(x)$ and $s_{n}(x)$ in each iteration can
slow down the numerical implementation of the AIM considerably and also lead
to precision problems. These drawbacks were overcome in the improved version
of AIM \cite{Cho:2010}. $\lambda _{n}(x) $ and $s_{n}(x)$ can be expanded in
Taylor series around the point $\xi $ at which the AIM is performed,%
\begin{equation}
\lambda _{n}(\xi )=\sum_{i=0}^{\infty }c_{n}^{i}(x-\xi )^{i},\text{ \ and \ }%
s_{n}(x)=\sum_{i=0}^{\infty }d_{n}^{i}(x-\xi )^{i}.  \label{eq:AIM_expansion}
\end{equation}%
Here $c_{n}^{i}$ and $d_{n}^{i}$ are the $i$th Taylor coefficients of $%
\lambda _{n}(\xi )$ and $s_{n}(\xi )$ respectively. Substituting these
expansions into Eq.~(\ref{eq:AIM_iteration}), we obtain a set of recursion
relations for the coefficients as 
\begin{equation}
c_{n}^{i}=(i+1)c_{n-1}^{i+1}+d_{n-1}^{i}+%
\sum_{k=0}^{i}c_{0}^{k}c_{n-1}^{i-k},\text{ \ and \ }%
d_{n}^{i}=(i+1)d_{n-1}^{i+1}+\sum_{k=0}^{i}d_{0}^{k}c_{n-1}^{i-k}.
\label{eq:AIM_exp_iteration}
\end{equation}%
Consequently, the ``quantization condition" can be expressed as%
\begin{equation}
d_{n}^{0}c_{n-1}^{0}-d_{n-1}^{0}c_{n}^{0}=0,  \label{eq:quantization}
\end{equation}%
which no longer requires the derivative operator. Both the accuracy and
speed of the AIM computation are greatly improved by this change.

Now we apply this method to charged BTZ black holes. The massless
Klein-Gordon equation is 
\begin{equation}
\nabla ^{\nu }\nabla _{\nu }\Psi =0.  \label{ChSc}
\end{equation}%
The solution of (\ref{ChSc}) can be considered in the form of%
\begin{equation}
\Psi =e^{-i\omega t}R(r)Y(\theta),
\end{equation}%
that makes us able to decompose the differential equation (\ref{ChSc}) into
two parts%
\begin{equation}
\frac{\partial ^{2}Y(\theta)}{\partial \theta^{2}}=0,
\end{equation}%
\begin{equation}
R^{\prime \prime }(r)+\left[ \frac{f^{\prime }(r)}{f(r)}+\frac{1}{r}\right]
R^{\prime }(r)+\frac{\omega ^{2}R(r)}{f(r)^{2}}=0.  \label{Rad}
\end{equation}%
where we set the separation constant to zero. Making definitions%
\begin{equation}
R(r)=\frac{\psi (r)}{\sqrt{r}},
\end{equation}%
and%
\begin{equation}
dr_{\ast }=\frac{dr}{f(r)},
\end{equation}%
where $r_{\ast }$ is the tortoise coordinate, we can rewrite (\ref{Rad}) in
the Schr\"{o}dinger form%
\begin{equation}
\frac{\partial ^{2}\psi }{\partial r_{\ast }^{2}}+\left( \omega
^{2}-V(r)\right) \psi =0,  \label{eq:radial_eq}
\end{equation}%
where%
\begin{equation}
V(r)=\frac{f(r)f^{\prime }(r)}{2r}-\frac{f(r)^{2}}{4r^{2}}.
\end{equation}%
We take a coordinate transformation $\xi =1-r_{+}/r$. At Infinity $%
r\rightarrow \infty $, $\xi \rightarrow 1$ and at horizon $r\rightarrow
r_{+} $, $\xi \rightarrow 0$ and therefore we can choose the point $\xi $
between $0$ and $1$. Applying $\xi =1-r_{+}/r$, Eq. (\ref{eq:radial_eq})
turns to%
\begin{equation}
\frac{\partial ^{2}\psi }{\partial \xi ^{2}}=\lambda _{0}(\xi )\frac{%
\partial \psi }{\partial \xi }+s_{0}(\xi )\psi .  \label{AIM}
\end{equation}%
We will express $\lambda _{0}$ and $s_{0}$ for linearly and nonlinearly
charged cases separately in the following of this section and apply AIM to
calculate the QNMs correspondingly to each case.

\subsection{Linearly charged case}

In the linearly charged case we have%
\begin{eqnarray}
\lambda _{0}(\xi ) &=&\frac{1}{1-\xi }\left[ 3+\frac{2ik\omega \left( 1-\xi
\right) }{\xi }\right] +\frac{2l^{2}\left( 1-\xi \right) ^{2}\left(
m-q^{2}+2q^{2}\ln {\frac{r_{+}}{l\left( 1-\xi \right) }}\right) }{%
l^{2}m\left( 1-\xi \right) ^{2}-r_{+}^{2}+2l^{2}q^{2}\left( 1-\xi \right)
^{2}\ln {\frac{r_{+}}{l\left( 1-\xi \right) }}},  \label{eq:lambda0} \\
s_{0}(\xi ) &=&-\frac{1}{4\left( 1-\xi \right) ^{4}}\left[ 3\left( 1-\xi
\right) ^{2}-\frac{12ik\omega \left( 1-\xi \right) ^{3}}{\xi }-\frac{%
4k\omega \left( k\omega -i\right) \left( 1-\xi \right) ^{4}}{\xi ^{2}}\right]
\notag \\
&&-\frac{1}{4\left( 1-\xi \right) ^{4}}\left[ \frac{4l^{2}\left( 1-\xi
\right) ^{4}\left( 2ik\omega +\xi \left( 3-2ik\omega \right) \right) \left(
m-q^{2}+2q^{2}\ln {\frac{r_{+}}{l\left( 1-\xi \right) }}\right) }{\xi \left(
l^{2}m\left( 1-\xi \right) ^{2}-r_{+}^{2}+2l^{2}q^{2}\left( 1-\xi \right)
^{2}\ln {\frac{r_{+}}{l\left( 1-\xi \right) }}\right) }\right]  \notag \\
&&-\frac{r_{+}^{2}}{\left( 1-\xi \right) ^{4}}\left[ \frac{\left( \omega
^{2}-\frac{1}{4l^{4}r_{+}^{2}\left( 1-\xi \right) ^{2}}\right) \left(
r_{+}^{2}-l^{2}m\left( 1-\xi \right) ^{2}-2l^{2}q^{2}\left( 1-\xi \right)
^{2}\ln {\frac{r_{+}}{l\left( 1-\xi \right) }}\right) }{\left( m-\frac{%
r_{+}^{2}}{l^{2}\left( 1-\xi \right) ^{2}}+2q^{2}\ln {\frac{r_{+}}{l\left(
1-\xi \right) }}\right) ^{2}}\right]  \notag \\
&&\times \left( 3r_{+}^{2}+l^{2}\left( m-4q^{2}\right) \left( 1-\xi \right)
^{2}+2l^{2}q^{2}\left( 1-\xi \right) ^{2}\ln {\frac{r_{+}}{l\left( 1-\xi
\right) }}\right) ,  \label{eq:s0}
\end{eqnarray}%
where $k=\left[ f^{\prime} \left(r_{+}\right)\right] ^{-1}$ and $\omega $ is
the quasi normal frequency. Substituting~(\ref{eq:lambda0}) and (\ref{eq:s0}%
) into the (\ref{AIM}), we can obtain the QNMs of charged BTZ black holes in
the presence of Maxwell electrodynamic field by using (\ref%
{eq:AIM_exp_iteration}) and (\ref{eq:quantization}). 
\begin{figure}[t]
\centering
\includegraphics[width=0.45\textwidth]{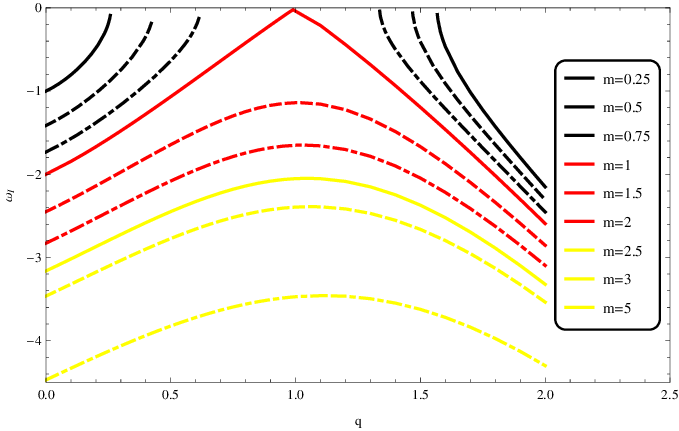}
\caption{{}The behavior of imaginary parts of QNMs ($\protect\omega _{I}$)
with respect to $q$. }
\label{fig:Modes-q-AIM}
\end{figure}

Calculations show that the real part of QNMs ($\omega _{R}$) is always zero.
So in the charged BTZ black hole, there is no oscillation of the dynamic
perturbation. This is consistent with the result found in the neutral BTZ
black hole \cite{Cardoso}. The dynamical perturbation only decays with the
relaxation time scale measured by the imaginary part of QNMs ($\omega _{I}$%
), whose dependence on the model parameters is shown in Fig. \ref%
{fig:Modes-q-AIM}. For the nonextreme black hole in the allowed parameters
range discussed above, the dynamical perturbation will die out which ensures
the stability of the black hole spacetime. But when the black hole
approaches the extreme case, $\omega _{I}$ becomes less negative and the
decay of the perturbation becomes slower. In the extreme case, $\omega _{I}$
becomes zero. This indicates that for the extreme black hole, the
perturbation will never die out which makes the black hole spacetime
unstable. For the cases $m<1$, $q\geq q_2$ and $m=1$, $q\geq 1$, with the
increase of $q$, we observe that the perturbation can decay faster, which
makes the black hole even more stable. When $m>1$, the black hole can never
become extreme and the $\omega _{I}$ is always negative.

It is interesting that the QNMs behavior exactly reflects what we discussed
for the spacetime property. Especially we find that when the black hole
phase transition happens, the dynamical perturbation presents us a
drastically different property. Nonextreme black hole can always be stable
with the decay of the dynamical perturbation, while the perturbation on the
extreme black hole background can persist which makes the extreme black hole
quasi-stable. This serves another character to mark the phase transition and
indicates different properties of different phases in the black hole.

\subsection{Nonlinearly charged case}

In nonlinearly BI charged case, we have 
\begin{eqnarray}
\lambda _{0}(\xi ) &=&\frac{3\xi +2ik\omega -2ik\omega \xi }{\xi -\xi ^{2}} 
\notag \\
&&+\frac{2\left( 1-\xi \right) \left( m-q^{2}+2q^{2}B(\xi )\right) }{\left(
m-q^{2}\right) \left( 1-2\xi +\xi ^{2}\right) -r_{+}^{2}\left( 2\beta
^{2}-1\right) +2r_{+}^{2}\beta ^{2}A(\xi )+2q^{2}\left( 1-\xi \right)
^{2}B(\xi )},  \label{eq:lambda0NL} \\
s_{0}(\xi ) &=&-\frac{1}{\left( 1-\xi \right) ^{2}}\left[ 1-\frac{3ik\omega
\left( 1-\xi \right) }{\xi }-\frac{k\omega \left( 1-\xi \right) ^{2}\left(
k\omega -i\right) }{\xi ^{2}}\right]  \notag \\
&&-\frac{\left( 2ik\omega +\xi \left( 3-2ik\omega \right) \right) \left(
m-q^{2}+2q^{2}B(\xi )\right) }{\xi \left[ \left( m-q^{2}\right) \left(
1-2\xi +\xi ^{2}\right) -r_{+}^{2}\left( 2\beta ^{2}-1\right)
+2r_{+}^{2}\beta ^{2}A(\xi )+2q^{2}\left( 1-\xi \right) ^{2}B(\xi )\right] }
\notag \\
&&-\frac{\omega ^{2}}{r_{+}^{2}\left( -1+2\beta ^{2}\left( -1+A(\xi )\right)
+\frac{m\left( 1-\xi \right) ^{2}}{r_{+}^{2}}+\frac{q^{2}\left( 1-\xi
\right) ^{2}}{r_{+}^{2}}\left( -1+2B(\xi )\right) \right) ^{2}}  \notag \\
&&+\frac{r_{+}^{2}\left( -1-2\beta ^{2}+2\beta ^{2}A(\xi )\right) }{\left(
1-\xi \right) ^{2}\left[ \left( m-q^{2}\right) \left( 1-2\xi +\xi
^{2}\right) -r_{+}^{2}\left( 2\beta ^{2}-1\right) +2r_{+}^{2}\beta ^{2}A(\xi
)+2q^{2}\left( 1-\xi \right) ^{2}B(\xi )\right] },  \notag \\
&&  \label{eq:s0NL}
\end{eqnarray}%
where 
\begin{equation}
A(\xi )=\sqrt{1+\frac{q^{2}\left( 1-\xi \right) ^{2}}{r_{+}^{2}\beta ^{2}}},%
\text{ \ and \ }B(\xi )=\ln {\frac{r_{+}\left( 1+A(\xi )\right) }{2\left(
1-\xi \right) }}.
\end{equation}%
Substituting~(\ref{eq:lambda0NL}) and (\ref{eq:s0NL}) into the (\ref{AIM}),
we can calculate the QNMs of BI charged BTZ black holes by using (\ref%
{eq:AIM_exp_iteration}) and (\ref{eq:quantization}).

\begin{figure}[t]
\centering%
\subfigure[~$m=0.25$]{
 \label{fig8a}\includegraphics[width=.3\textwidth]{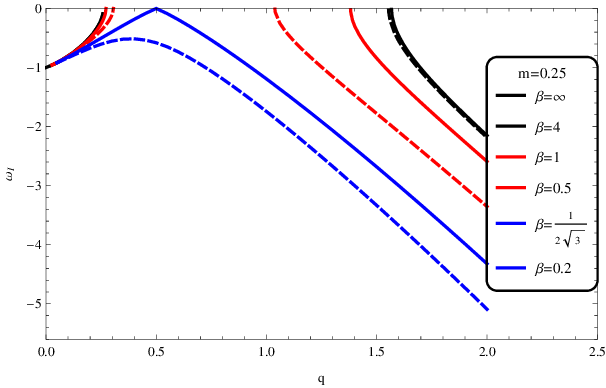} } 
\subfigure[~$m=1$]{
 \label{fig8b}\includegraphics[width=.3\textwidth]{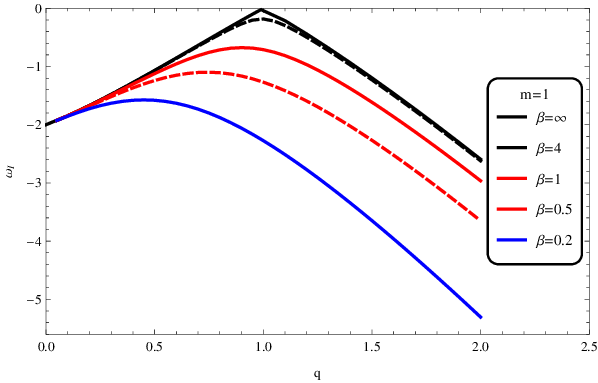} } 
\subfigure[~$m=2$]{
 \label{fig8c}\includegraphics[width=.3\textwidth]{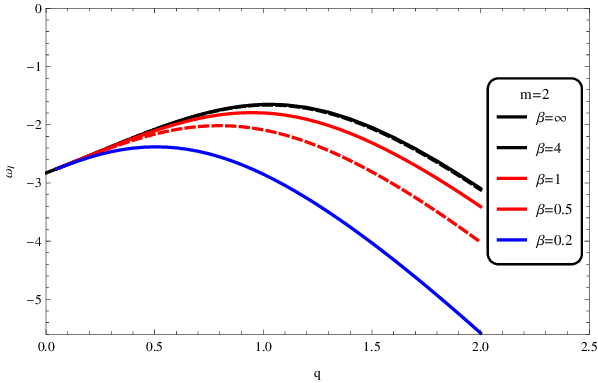} }
\caption{{}The behavior of imaginary parts of QNMs ($\protect\omega _{I}$)
with respect to $q$ for nonlinearly charged case.}
\label{fig:Modes-q-AIM025NL}
\end{figure}

Again the real part of QNMs ($\omega _{R}$) vanishes indicating that there
is no oscillation of the perturbation. The decay time scale of the
perturbation can be measured by the imaginary part of $\omega _{I}$ which is
shown in Figure \ref{fig:Modes-q-AIM025NL}. This behavior is similar to that
of the linearly charged case.

For the nonextreme black hole, the spacetime is always stable, since the
dynamic perturbation will die out finally. When the nonextreme black hole
approaches extreme with the increase of the electric charge, the
perturbation will die out more slowly, since the $\omega _{I}$ becomes less
negative. For the extreme black hole, $\omega _{I}$ is zero which indicates
that for the extreme black hole background, the perturbation will persist
and will not die out. Compared with the linearly charged BTZ black hole
case, we have observed that the nonlinearity introduced here helps to
protect the stability of the black hole. Increasing of the nonlinearity, the 
$\omega _{I}$ becomes more negative for the same parameters $m$ and $q$
case. This agrees with the observation in the four-dimensional BI AdS black
hole \cite{Wang5}. The QNMs behavior reflects the spacetime properties
discussed above. Interestingly we again witnessed that the dynamic
perturbation can be a signature to probe the phase transition in the system.
We see the evidence that two phases, the nonextreme and extreme black holes
have different dynamical behaviors. This actually supports the phase
transition results obtained in the previous section.

\section{Summary and conclusions}

In this paper, we have carefully examined the spacetime properties of BTZ
black holes in Maxwell field and Born-Infeld field. We found new, rich
spacetime structures of the black hole when the model parameters are varied.
These special spacetime properties determine the conditions of phase
transition. For a charged BTZ black hole with mass in a small value regime, $%
0<m<1$, the black hole can evolve from nonextreme to extreme black hole when
the charge of the black hole increases from zero to some $q_{1}$. When the
charge of the black hole increases further, there will be no black hole
solution. But when the charge parameter becomes as big as some $q_{2}$, an
extreme black hole appears and this extreme black hole will become
nonextreme with the further increase of the charge parameter. The difference
between $q_{1}$ and $q_{2}$ will become smaller with the increase of $m$ and 
$q_{1}$, $q_{2}$ will degenerate when $m=1$. But when the mass parameter is
bigger than 1, the black hole will always be nonextreme no matter how much
we increase the charge parameter. Generalizing the discussion to the
nonlinear charged BTZ black holes, we find the qualitative behaviors in the
structures of black hole persist when the BTZ black hole with nonlinear
Born-Infeld electromagnetic field.

Furthermore we have studied the thermodynamical phase transition in the
charged BTZ black hole background. Instead of the heat capacity, we have
employed the Landau-Lifshitz theory and examined the second moments of
relevant parameters in the system. We have found that some second moments
diverge when the black hole extreme limit is reached, which indicates the
break down of the rigorous meaning of thermodynamical quantities. This marks
the occurrence of the thermodynamical phase transition, where some physical
properties change.


To disclose more signature of the phase transition, we have calculated the
QNMs for the charged BTZ black holes. The dynamical perturbation does not
oscillate in the charged BTZ black hole. It only shows the decay behavior
when the charged BTZ black hole is nonextreme, which shows that the
perturbation outside the nonextreme black hole will finally die out so that
the nonextreme charged BTZ black hole is always stable. When the nonextreme
black hole evolves towards the extreme hole, the decay becomes slower. But
when the extreme black hole is reached, the dynamical perturbation will not
decay, it persists. This is a dangerous signal which tells us that the
extreme charged BTZ black hole is easily destroyed if we add more
perturbations to the system. Considering the nonlinearity of the
electromagnetic field, we have observed that the nonlinearity can partially
save the black hole and make the perturbation outside the black hole decay
faster. The different properties of QNMs for the extreme and nonextreme
charged BTZ black holes are interesting. They can serve as another probe for
the phase transition between the nonextreme and extreme black holes.

We have discussed the relation between horizons and found that there exists
a turning value in the mass parameter, $m_{r}$. When $m>1$, the black hole
is always nonextreme, but in the range $1<m<m_{r}$, two horizons can first
approach each other and then separate away with the increase of charge $q$,
but when $m>m_{r}$, the difference between two horizons will become bigger
monotonically when $q$ increases. At first sight, this property looks
similar to the temperature behavior, since when $m>1$, black hole
temperature also has a turning point at $m_{t}$. But we found that $m_{r}$
and $m_{t}$ do not coincide. The reason can be understood that temperature
is just an apparent thermodynamical quantity, it cannot reflect deep physics
in the spacetime property. In the QNM behavior, we found that when the black
hole horizons get closer, the perturbation around the black hole decays
slower and slower. But when the difference between black hole horizons
becomes bigger, the black hole becomes more nonextreme, the perturbation
around the black hole decays faster. The value $m_{r}$ is really the
boundary parameter we can distinguish the speed of the decay of
perturbations around black holes. However this behavior change in the QNMs
does not show up in $m_{t}$. This tells us that the black hole temperature
is not a good index to reflect the spacetime dynamical property.

The zero temperature of the extreme black hole is a thermodynamical
indication of the occurrence of a second order phase transition between
nonextreme and extreme black holes \cite{Pavon}. But this is not a direct
reason for the behaviour of QNMs. The vanishing decay of the dynamical
perturbation is an intrinsic result of the phase transition happening
between nonextreme and extreme black holes. It is not caused directly by the
zero temperature of the black hole.

\begin{acknowledgments}
This work was supported in part by NNSF of China. MKZ would like to thank
Shanghai Jiao Tong University for the warm hospitality during his visit.
This work has been supported financially by Research Institute for Astronomy
\& Astrophysics of Maragha (RIAAM). We thank Yen Chin Ong for useful
discussions.
\end{acknowledgments}

\end{document}